\documentclass[usenatbib]{mnras}

\usepackage[T1]{fontenc}
\usepackage{ae,aecompl}

\usepackage{graphicx}	
\usepackage{amssymb}
\usepackage[fleqn]{amsmath}

\newcommand{\kms}{\mbox{km\,s$^{-1}$}}
\newcommand{\jupm}{M$_{\rm{J}}$}
\newcommand{\bsig}{\mbox{BANYAN $\Sigma$}}

\begin{document}

\title[The Age of the Carina Association]{The Age of the Carina Young Association and Potential Membership of HD~95086}
\author[M. Booth, C. del Burgo and V. V. Hambaryan]{Mark Booth$^{1}$\thanks{E-mail: markbooth@cantab.net}, Carlos del Burgo$^{2,3}$ and Valeri V. Hambaryan$^{1,4}$\\
$^{1}$Astrophysikalisches Institut und Universit\"atssternwarte, Friedrich-Schiller-Universit\"at Jena,
      Schillerg\"a{\ss}chen~2--3, 07745 Jena, \\Germany \\
$^{2}$Instituto Nacional de Astrof\'isica, \'Optica y Electr\'onica, Luis Enrique Erro 1, Sta. Ma. Tonantzintla, Puebla, Mexico \\
$^{3}$Instituto de Astrof\'isica de Canarias, Calle V\'ia L\'actea S/N 38205 La Laguna, Tenerife, Spain \\
$^{4}$Byurakan Astrophysical Observatory, Byurakan 0213, Aragatzotn, Armenia
}

\date{Accepted 2020 November 13. Received 2020 November 13; in original form 2020 July 17}

\maketitle

\begin{abstract}
Carina is a nearby young stellar association. So far, only a small number of stars have been clearly identified as members of this association.
In this paper we reanalyse the membership of the association in light of \emph{Gaia} DR2 data, in particular finding that HD~95086 is a potential member (probability of 71\%). This star is noteworthy as one of the few stars that hosts both a detected debris disc and a directly imaged planet. It has previously only been considered as a potential member of the Lower Centaurus Crux (LCC) -- part of the Scorpius-Centaurus association. 
We also reanalyse the age of the Carina association. Using a Bayesian inference code applied to infer a solution from stellar evolution models for the most probable (>99\%) members of Carina, we infer an age for the association of 13.3$^{+1.1}_{-0.6}$~Myr, much younger than previous studies. Whilst we have revised HD~95086's association membership from LCC to Carina, the fact that we also find Carina to have a younger age, similar to that of LCC, means that the estimates of HD~95086b's mass remain unchanged. However, the younger age of Carina does mean that the companion to another Carina member, HD~44627 (or AB Pic), has a mass that is more clearly in the planet rather than brown dwarf range.

\end{abstract}


\section{Introduction}

There are a number of methods for determining the age of a star. These include using gyrochronology, X-ray emission, lithium depletion, chromospheric activity and stellar evolution models. Not all methods work equally well for all spectral types and evolution phases, and the accuracy for an individual star is usually not high \citep[see e.g.][]{vican12}.

However, most stars are thought to form in clusters that slowly disperse into associations \citep[e.g.][]{ambartsumian47, blaauw64} before dispersing completely into the Galactic field. When an association of stars is identified, then its age can be determined more accurately by assuming all the stars are coeval.

Based on a survey of X-ray sources in the southern hemisphere and kinematic information from \emph{Hipparcos}, \citet{torres01} identified the existence of the Great Austral Young Association. Further study of this association by \citet{torres08}, demonstrated that it should be subdivided into at least three young associations: Tucana-Horologium (Tuc-Hor), Columba and Carina. These are expected to be related and likely of a similar age, but are distinguished by their velocity components. Whilst Tuc-Hor is compact and well defined, the Columba and Carina associations are more diffuse and their membership is debated. This is particularly true of Carina, with, for example, only 7 bona fide members listed by \citet{gagne18} compared to 20 and 45 bona fide members for Columba and Tuc-Hor respectively.

In this paper we reconsider the membership of the Carina association in \autoref{scmems} based on the \emph{Gaia} DR2 data. In doing so, we propose that HD~95086 is a potential Carina member for the first time. The age of this star is of particular interest as it hosts both a directly imaged planet and a resolved debris disc. In \autoref{scage} we then reconsider the age of the association, first by comparing stellar evolution models to the \emph{Gaia} photometry and then by considering the kinematic evolution of the association. We conclude in \autoref{sdiscuss} by discussing the implications of these findings.

\begin{table*}
\begin{tabular}{lrllr}
                     Name &     \textit{Gaia} DR2 ID &      $v_r$ km$,s^{-1}$ &         $v_r$ source &   Prob. (\%) \\ 
\hline
HD 49855 &  5265670762922792960 &  20.48$\pm$0.27 &        \citep{gaia18} &      100.0 \\
           UCAC3 53-40215 &  5297100607744079872 &  21.17$\pm$0.41 &   \citep{schneider19} &      100.0 \\
  2MASS J08040534-6316396 &  5277462269217606400 &   22.0$\pm$0.85 &   \citep{schneider19} &      100.0 \\
  2MASS J09315840-6209258 &  5250988846733325312 &  19.21$\pm$1.24 &   \citep{schneider19} &       99.9 \\
                 HD 42270 &  4621305817457618176 &    16.7$\pm$0.6 &      \citep{torres06} &       99.9 \\
  2MASS J08063608-7444249 &  5213934514587912320 &   17.0$\pm$1.04 &        \citep{gaia18} &       99.7 \\
                 HD 37402 &  4759444786175885824 &    23.7$\pm$0.5 &  \citep{gontcharov06} &       99.7 \\
                HD 298936 &  5356713413789909632 &   18.4$\pm$0.44 &        \citep{gaia18} &       99.6 \\
                    m Car &  5251098523021221376 &    20.0$\pm$4.0 &  \citep{gontcharov06} &       99.6 \\
  2MASS J06262199-7516404 &  5261554496328279552 &  16.45$\pm$2.18 &        \citep{gaia18} &       99.4 \\
                 V479 Car &  5299141546145254528 &  19.26$\pm$0.92 &        \citep{gaia18} &       99.3 \\
                 HD 44627 &  5495052596695570816 &  21.86$\pm$0.37 &        \citep{gaia18} &       98.9 \\
                 HD 55279 &  5208216951043609216 &  16.44$\pm$0.45 &        \citep{gaia18} &       98.7 \\
                   AL 442 &  5266182443853174784 &  20.96$\pm$1.15 &   \citep{schneider19} &       97.7 \\
  2MASS J07441105-6458052 &  5287415735666649728 &    19.9$\pm$2.7 &        \citep{gaia18} &       96.9 \\
  2MASS J09180165-5452332 &  5310606287025187456 &  25.43$\pm$3.12 &   \citep{schneider19} &       96.9 \\
  2MASS J07065772-5353463 &  5491506843495850240 &   22.57$\pm$0.9 &   \citep{schneider19} &       96.2 \\
 WISE J080822.18-644357.3 &  5277096097486882560 &    22.7$\pm$0.5 &      \citep{murphy18} &       95.1 \\
  2MASS J07013884-6236059 &  5286760525517037568 &  22.57$\pm$0.72 &   \citep{schneider19} &       94.6 \\
           TYC 8602-718-1 &  5308164516541067648 &   22.65$\pm$1.4 &        \citep{gaia18} &       94.4 \\
                 HD 83096 &  5217846851839896704 &  22.44$\pm$0.78 &        \citep{gaia18} &       93.6 \\
                  iot Hyi &  4626843786944938880 &   14.6$\pm$0.21 &        \citep{gaia18} &       89.6 \\
  2MASS J04082685-7844471 &  4625883599760005760 &  14.25$\pm$0.86 &        \citep{gaia18} &       86.3 \\
  2MASS J08194309-7401232 &  5219983787046519168 &  26.24$\pm$1.54 &   \citep{schneider19} &       82.7 \\
                 HD 95086 &  5231963962676292224 &    17.0$\pm$2.0 &        \citep{moor13} &       70.8 \\
           TYC 9200-446-1 &  5219351911459314048 &  18.01$\pm$1.48 &        \citep{gaia18} &       57.5 \\
\end{tabular}
\caption{HD~95086 and potential members of Carina (as identified in \autoref{scmems} and ordered by their membership probability) with their radial velocities, radial velocity source and probability of membership as determined by \bsig.}
\label{tcarmems}
\end{table*}

\section{Carina Membership}
\label{scmems}
Stellar associations were first determined by finding stars that are close together on the sky, are at a similar distance and moving in a similar direction. For nearby associations, the stars typically spread across the sky and can be mistaken for foreground stars, so it is necessary to consider their Galactic coordinates, $XYZ$, and space velocities, $UVW$ \citep[e.g. see the reviews by][]{zuckerman04a,torres08}. Various methods have been developed to do this, which are described in detail by \citet{riedel17}. For our purposes we shall use the \bsig{} code \citep{gagne18} to determine association membership as this is the most comprehensive code available, including 27 young associations out to a distance of 150~pc. BANYAN (Bayesian Analysis for Nearby Young AssociatioNs) is an algorithm based on Bayesian inference, where associations are modeled with unidimensional Gaussian distributions in Galactic coordinates $XYZ$ and space velocities $UVW$ originally developed by \citet{malo13}. This was further improved to model associations as freely rotating three-dimensional Gaussian ellipsoids in position and velocity space for BANYAN II \citep{gagne14}. \bsig{} works with updated and new models of young associations and uses full six-dimensional multivariate Gaussians in position and velocity space. In order to use this code, a minimum of the celestial coordinates and proper motions are required, but parallaxes and radial velocities (RVs) can also be provided to further constrain the results.

\citet{malo13,moor13b,malo14,elliott15,silverberg16,gagne18,gagne18a,gagne18b,schneider19} have all suggested various stars as members of Carina. We compile an initial list from these papers, but first need to determine the membership probability using the latest position and velocity data. In order to do this we cross match the stellar coordinates of each of the potential members with the \emph{Gaia} database to determine their \textit{Gaia} DR2 ID and retrieve their coordinates, parallaxes, proper motions and (where available) RVs. 

As \textit{Gaia} DR2 does not include RVs for all stars, for a number of stars we used RVs from an alternative source. Note that in the case of spectroscopic binaries, care must be taken as the RV can vary strongly as stars orbit each other. In our sample, two stars are listed in the Washington Double Star Catalog \citep{mason01}: HD~83096 and AL~442. The RV for HD~83096 was measured by \emph{Gaia} over 14 transits so any deviations due to the binary companion will have been averaged out. The RV for AL~442 was not measured by \emph{Gaia}, however, in addition to the one measurement by \citet{schneider19} that we use in our calculations, two more measurements were made by \citet{durkan18}: $v_r=19.33\pm0.29$ ~km$,s^{-1}$ and $v_r=19.54\pm0.26$ ~km$,s^{-1}$. Neither of these is far enough from the \citet{schneider19} measurement to affect our results.

The six-dimensional position and velocity information was then passed to \bsig{} to check the association membership probabilities of these stars. The stars that are found to have a membership probability $>$50\% are presented in \autoref{tcarmems} along with their source IDs and radial velocities.

\begin{figure*}
	\centering
	\includegraphics[width=0.98\textwidth]{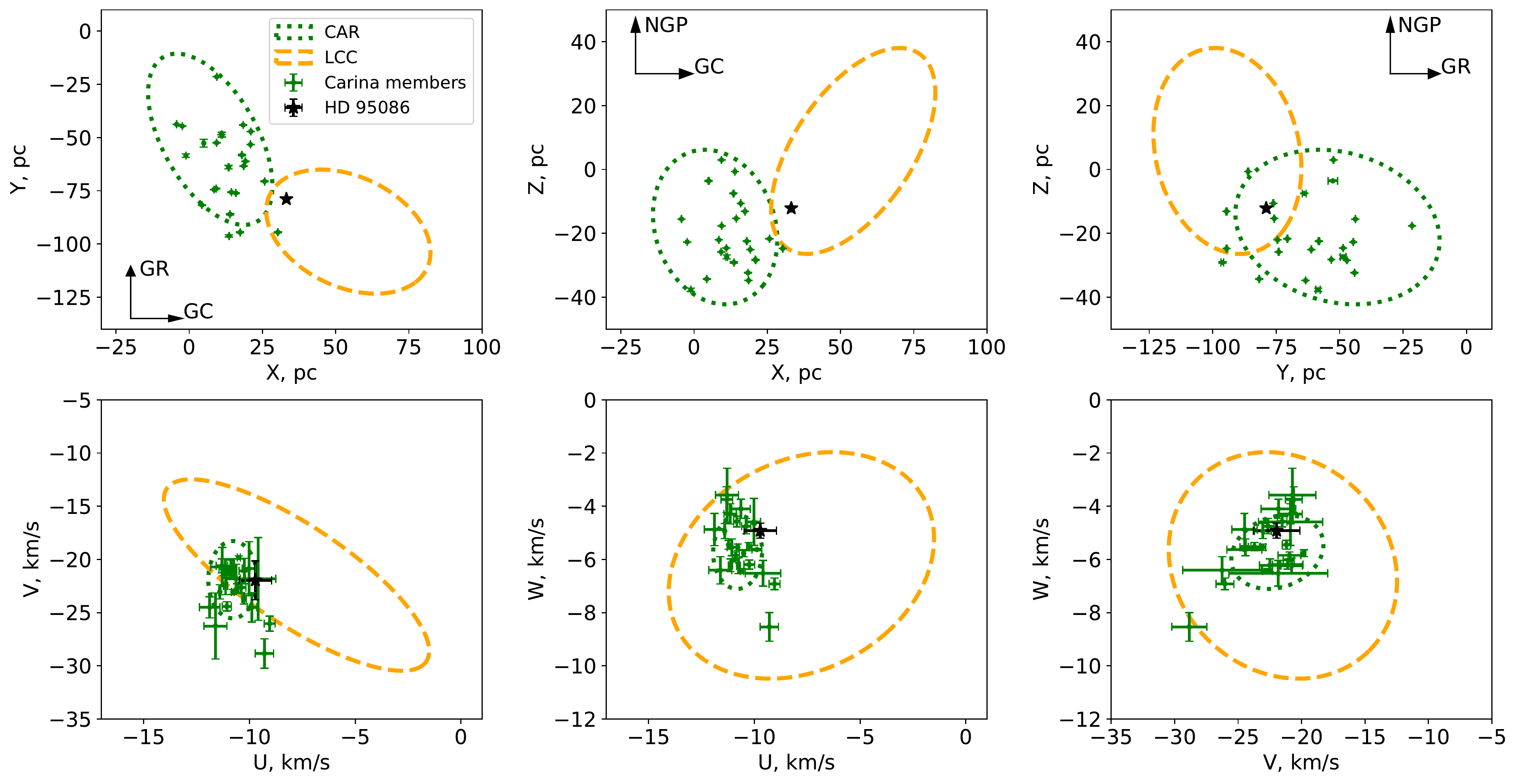}
	\caption{Plot of the position and velocity in Cartesian coordinates of the Carina (orange dotted ellipse) and LCC (green dashed ellipse) associations as well as HD~95086 (black point) and the rest of the potential Carina members described in \autoref{scmems} (green points). The ellipses represent the FWHM of the six-dimensional Gaussian models for the associations used by \bsig. The arrows point in the directions of the Galactic centre (GC), Galactic rotation (GR) and North Galactic Pole (NGP). HD~95086 is seen to lie within the FWHM of the LCC model for all coordinates, but its proximity to the centre of the Carina model, particularly in terms of velocities, increases the likelihood that it is a Carina member.}
	\label{fuvw}
\end{figure*}

\subsection{HD~95086}
\label{skin}
HD 95086 is a pre-main sequence star of spectral type A8 at a distance of 86.2$\pm$0.3~pc \citep{bailer18}. It is one of the few systems with both a directly imaged planet \citep{rameau13} and a debris disc \citep{chen12}. 
Precisely determining the age is necessary to infer the mass of the planet. It also helps us place the system in the context of the collisional evolution of debris discs \citep[e.g.][]{krivov18a,krivov18b}.

\citet{dezeeuw99} proposed that the star could be part of the Lower Centaurus Crux (LCC) -- a sub-region of the Scorpius-Centaurus Association -- based on \emph{Hipparcos} data, although they only gave it a probability of 41\%. \citet{rizzuto12} also included it as a member of LCC but with a higher probability of 81\%. In their case, this high value is because they interpreted the presence of a debris disc as a proxy for youth and, thus, increased the prior. \citet{damiani19} included it in the diffuse population of LCC, although they noted that a 10-30\% contamination rate due to the method used is expected. So far, however, no publication has considered the possibility that it is a member of a different association. Due to questions about its RV, particularly in relation to the possible presence of CO gas in its debris disc \citep{booth19}, we have reason to question its association membership.

\begin{figure*}
	\centering
	\includegraphics[width=0.48\textwidth]{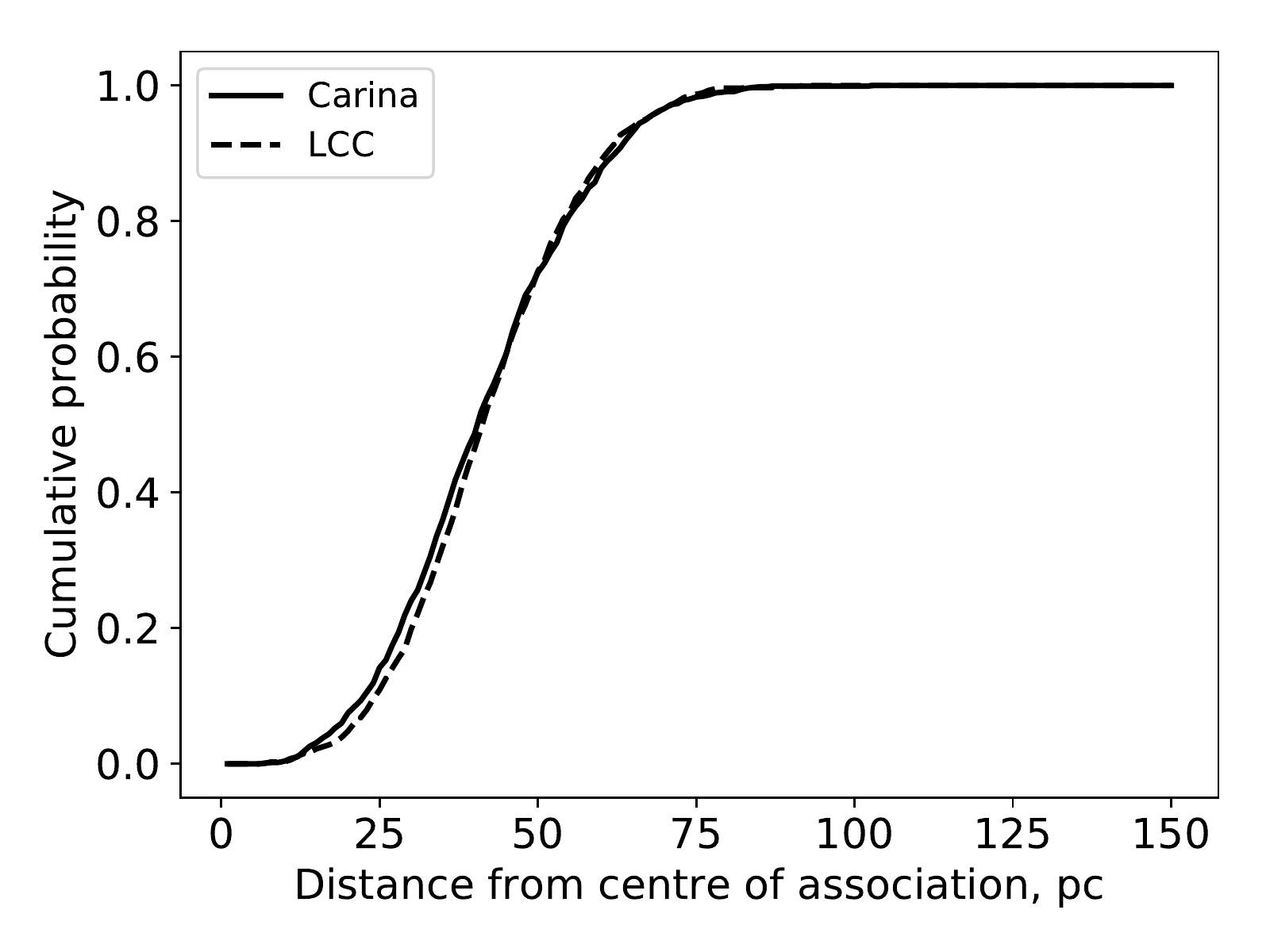}
	\includegraphics[width=0.48\textwidth]{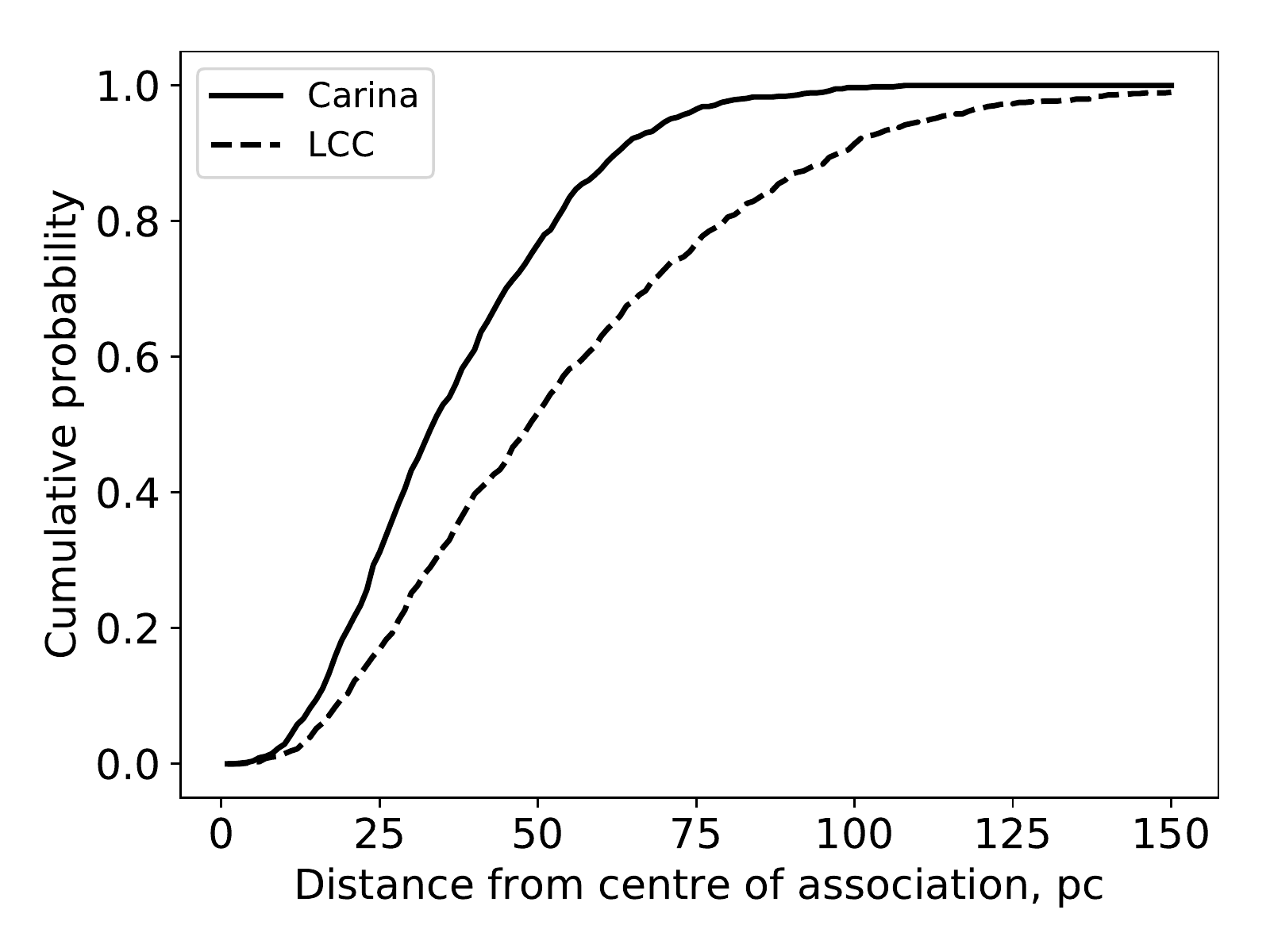}
	\caption{The cumulative distribution function of the distance between HD~95086 and the centre of the young associations Carina (solid line) and LCC (dashed line). \emph{Left: } The situation at the present time. From this it is seen that the star is not significantly closer to the centre of either association. \emph{Right: } The situation after tracing back for 10~Myr. By this point HD~95086 is clearly more likely to be closer to the centre of Carina than the centre of LCC.}
	\label{fcarvslcc}
\end{figure*}

As HD~95086 has an effective temperature $>$7000~K (the exact temperature is inferred along with the age in \autoref{scarage} and can be found in \autoref{tsprops}), its RV was not determined for the \emph{Gaia} DR2, for which RVs could only be accurately determined for stars in the effective temperature range 3500-7000~K \citep{sartoretti18}. However, two values for the RV of HD~95086 do exist in the literature. \citet{madsen02} used the \emph{Hipparcos} data together with the assumption that it is a member of LCC to estimate that it should have an RV of 10.1$\pm$1.2~\kms{}. In a similar manner we can use the \emph{Gaia} DR2 data with Banyan $\Sigma$ to revise this prediction of the expected RV. We find that the probability of HD~95086 being in LCC is 33.8\% and, if it is in this association, it should have an RV of 15.9$\pm$3.7 \kms, which is higher than the value from \citet{madsen02}. In addition, there is a probability of 48.5\% that the star is actually a member of the Carina association and, in that case, the RV would be 17.1$\pm$0.6 \kms{}.

\citet{moor13} spectroscopically measured the RV\footnote{The spectroscopic catalogue of \citet{kharchenko07} also includes HD~95086 and lists the radial velocity as 10.1$\pm$1.2~\kms{}, however there is an issue with this as described further in appendix \ref{skhar}.} to be 17$\pm$2 \kms{}. This perfectly matches the value expected if this is a Carina member and so, when we include this value of the RV as an input when running \bsig{} we find that the probability of this star being in LCC drops to 23.8\% whereas the probability of it being in Carina rises to 70.8\%. This is further illustrated in \autoref{fuvw}, which shows HD~95086 on the outskirts of Carina with regards to its position and with a velocity vector that is very close to that of the Carina model.

In addition to checking how the present day position and velocity of HD~95086 compares to the known young associations, we can also trace back the star's position and association distributions to determine how close they were in the past. 
To study the Galactocentric motion of HD~95086, Carina and LCC we make use of the code
described in \citet{neuhauser19}, which computes the orbits by a 
numerical integration
of their equations of motion as defined by the Galaxy gravitational potential
consisting of a three component (bulge, disk and halo) axisymmetric model
\citep[Model III from][]{bajkova17}. In addition, the Galaxy
gravitational potential is supplemented with the more realistic,
non-axisymmetric and time dependent terms, which take into account the
influence of the central bar and the spiral density wave 
\citep{palous93,fernandez08,bajkova19}.

In order to take account of the uncertainties in the astrometric parameters of the star and associations, each one was replaced by 1000 clones, each with astrometric parameters drawn from a multivariate normal distribution. This is done making use of the covariance matrix of astrometric parameters from \emph{Gaia} DR2 \citep{gaia18} for the star and from the \bsig{} association models for the associations \citep{gagne18}. 
Such a procedure is superior to the individual,
independent random drawing of each parameter that ignores their 
mutual dependence. 

From this process we have probability distribution functions of the separation distance between HD~95086 and the centre of the two associations. By comparing the cumulative distribution functions for the two associations we can determine which association HD~95086 was more likely closer to the centre of in the past. This is shown in \autoref{fcarvslcc}. From this we find that although HD~95086 is currently approximately equally close to the centre of both associations, tracing back in time to 10~Myr, we find that the likelihood of HD~95086 being closer to Carina was greater by a factor of between 2 and 4. This greatly strengthens our conclusion of HD~95086 being a likely Carina member as long as it is $>$10~Myr old. The ages of HD~95086 and Carina are assessed in the following two sections.

The main caveat here is that we have relied on the association models included as part of the \bsig{} algorithm. Recent research by \citet{lee19} implements a new algorithm for determining association models. They find that the parameters for most associations agree with those of \citet{gagne18}, but they do find significant differences for Carina. Their code is not public, but they do provide some details of the association models. The model of \citet{gagne18} locates the centre of the Carina association at $XYZ=(7.2, -50.8, 18.0)$~pc and $UVW=(-10.8, -21.9, -5.7)$~\kms{}, whereas the model of \citet{lee19} locates the centre of the Carina association at $XYZ=(12.3, -114.7, -18.3)$~pc and $UVW=(-10.6, -22.5, -4)$~\kms{}, i.e. there is a large shift in $Y$ but the other coordinates are not greatly changed. Their work does not include associations beyond 100~pc, such as LCC, and it is unclear how this might effect their results. Nonetheless, we can estimate how much using their model of Carina rather than that of \citet{gagne18} would affect our results. We find that the distance from HD~95086 to the centre of Carina is not significantly changed by this, being 39~pc for the former model and 42~pc for the latter and so we do not expect it to impact our conclusions about the membership of HD~95086.

\section{Age analysis of the Carina Association}
\label{scage}
The age of Carina has previously been estimated as 45$^{+11}_{-7}$~Myr \citep{bell15} based on isochrones and $>$28~Myr based on kinematics \citep{miret18}. However, \citet{schneider19} find that Carina has an age closer to that of the $\beta$ Pic Moving Group based on Lithium measurements. We, therefore, consider it worthwhile to re-evaluate the age of Carina.

\begin{figure}
	\centering
	\includegraphics[width=0.48\textwidth]{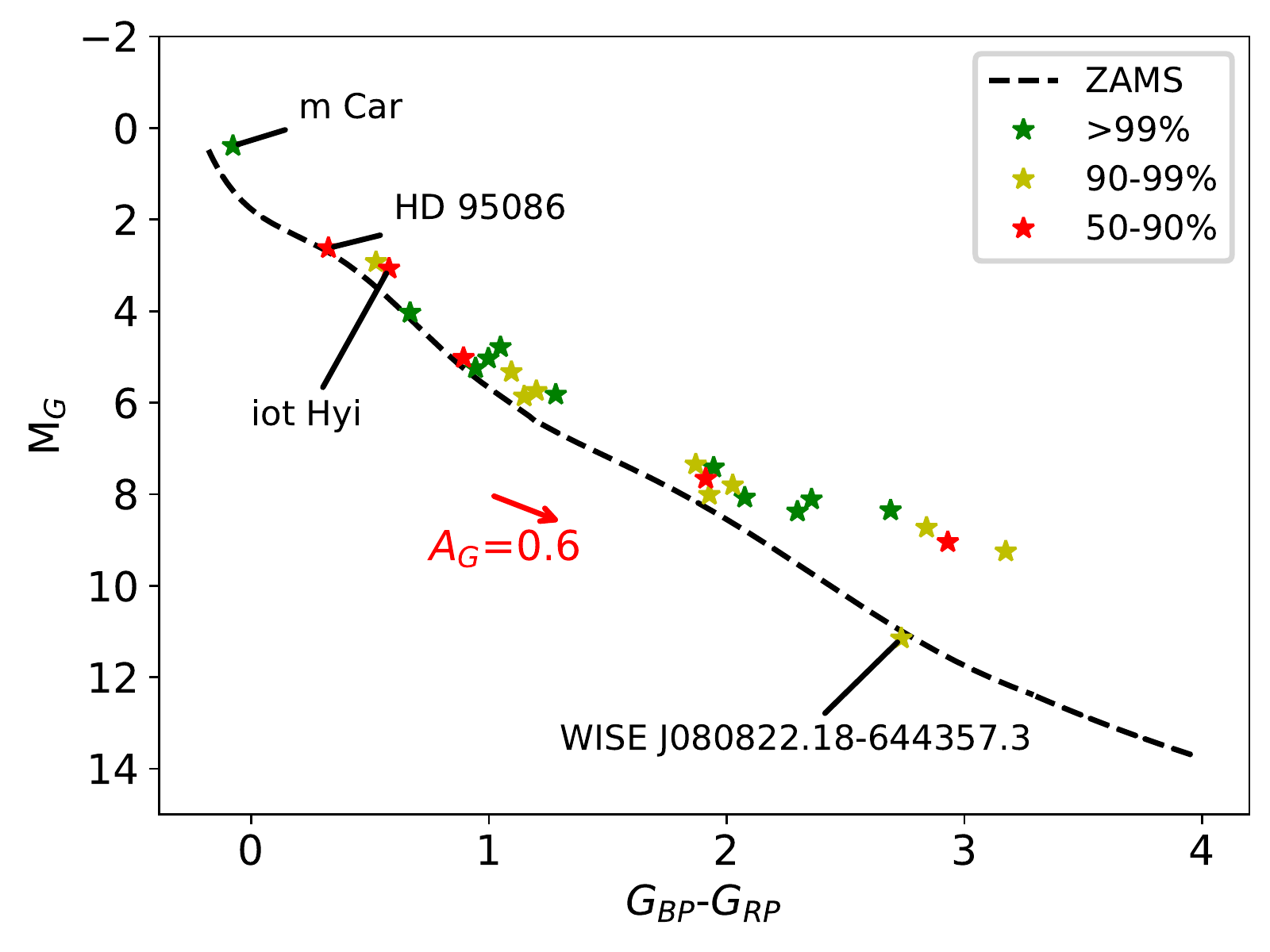}
	\caption{M$_G$, the absolute magnitude in the $G$ band, versus $G_{BP}$-$G_{RP}$ colour for Carina members identified in this paper (shown with star symbols, where the colour signifies the membership probability). The dashed line shows the zero age main sequence for [Fe/H]=0. The red arrow shows the average extinction vector for the sample.} 
	\label{fcamd}
\end{figure}

\subsection{The age of Carina inferred from stellar evolution models}
\label{scarage}
\autoref{fcamd} shows the colour magnitude diagram (CMD) for the stars that we have identified as potential members of the Carina association. For reference, 

we also plot the zero age main sequence (ZAMS) line, assuming [Fe/H]=0. We see that the majority of the stars lie close to ZAMS except for most of the M stars, which are above it. Offsets from the ZAMS line can be explained by a combination of the stars still being in the pre-main sequence phase, reddening due to the extinction and (in some cases) binarity. The effect of extinction has been estimated for some of the stars in our sample, with an average of $A_G=0.6$ and $E$(BP-RP)$=0.3$ (shown by the red arrow in \autoref{fcamd}). This is shown here for illustration and not corrected for as the extinction and redenning estimates provided by \emph{Gaia} are not expected to be accurate for individual stars \citep{andrae18} and the sample is too diverse (e.g. distances to our stars range from 30 to 100~pc) for the average to be an accurate estimate for individual stars.

HD~95086 has previously been identified as a giant star with \citet{houk75} giving it the classification A8III. However, as seen here and already noted by \citet{rameau13}, its position on the CMD clearly places it close to the ZAMS line with a temperature and luminosity (see \autoref{tsprops}) that are consistent with those expected for an A8V star \citep[e.g.][]{pecaut13,eker18}. Similarly, m Car and iot Hyi are both listed as having a spectral classification of IV-V by \citet{houk75} and \citet{gray06} respectively, yet both are only slightly offset from ZAMS and their offsets can be explained by a combination of extinction and being in the pre-main sequence phase of their evolution, so they are expected to be too young to have evolved off the main sequence.

Here we take a renewed look at the age inferred from stellar evolution models using the Bayesian inference code of \citet{delburgo18} and incorporating the latest data from \emph{Gaia} DR2 \citep{gaia18}. This Bayesian analysis makes use of the PARSEC v1.2S library of stellar evolution models \citep{bressan12,chen14,chen15,tang14}. It takes as input the absolute $G$ magnitude (obtained from the apparent magnitude and the parallax), $G_{BP}$-$G_{RP}$ colour, [Fe/H] and their uncertainties, returning theoretical predictions for other stellar parameters. We have assumed solar metallicity with [Fe/H]=0.00$\pm$0.20 and adopted the Gaia photometry passbands ($G$, $G_{BP}$, and $G_{RP}$) from \citet{maiz18}. As a prior, we assume that all of our stars are in the pre-main sequence phase due to the expected youth of the young association.

\begin{figure}
	\centering
	\includegraphics[width=0.5\textwidth]{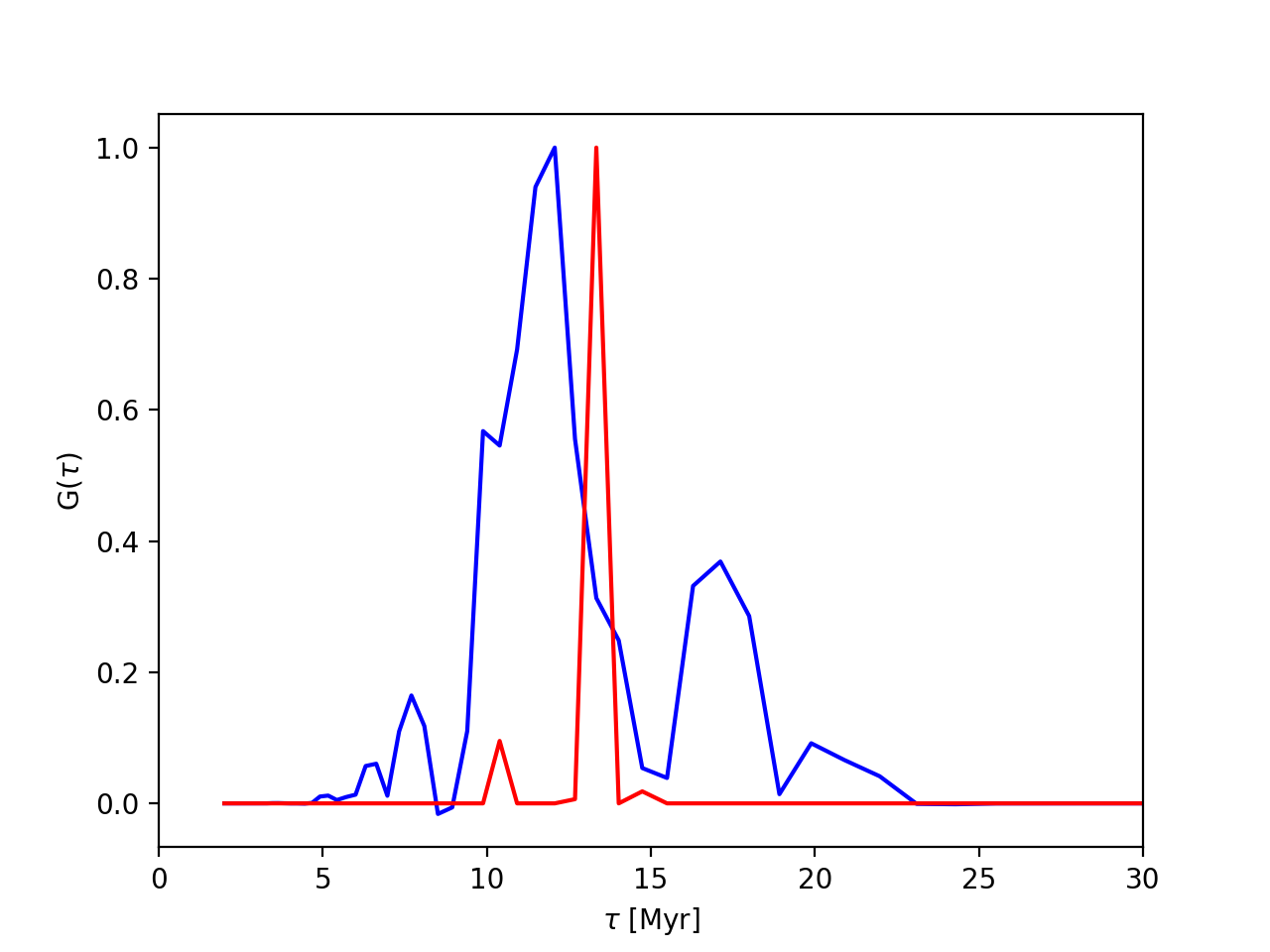}
	\caption{Posterior distribution for the age of the Carina association (red line) and HD~95086 (blue line). 
	}
	\label{fpostage}
\end{figure}

We determine ages for all of the stars in our membership list (\autoref{tcarmems}). We then multiply the posterior likelihoods of the individual stars to determine the age of the association under the assumption that the members are coeval. We first consider only the most likely members (the 11 stars with a $>$99\% probability of being in Carina) in order to have greater confidence that our age is not biased by stars that may not actually be members. We find that m Car, the only B star in the list, is just 3.3$\pm$0.4~Myr and so is excluded from this calculation as a clear outlier. Combining the posterior likelihoods for the other 10 stars in this group results in a posterior distribution that is shown in \autoref{fpostage}. The mode and 68\% uncertainties of this distribution are 13.3$^{+1.1}_{-0.6}$~Myr. This is much lower than pre-\emph{Gaia} estimates due to the accuracy of the \emph{Gaia} data used, both for the selection of Carina members and the age estimation. 
If we expand our sample for this calculation to include all stars with a membership probability $>$90\% (19 objects in total\footnote{Note that WISE J080822.18-644357.3 is not included here. 

\citep{murphy18} show that the PARSEC isochrones do not give a good fit for this particular star. For a more detailed analyis of this star's age we refer the interested reader to their paper.}) then we find an age of 14.7$^{+0.8}_{-0.7}$~Myr for the association, consistent with the age for just the most probable members, although we consider the age for just the most probable members to be the more reliable value.

As noted in \autoref{skin}, \citet{lee19} derived a model for the Carina association that is different to the one used in this paper from \citet{gagne18}, primarily in terms of its Galactic $Y$ coordinate. If the model from \citet{lee19} is correct, then some of the stars that we have considered as Carina members in this paper may not be Carina members after all. However, we do not find any correlation between a star's $Y$ coordinate and its age and so would not expect this to have a strong effect on the age of the Carina association derived in this section. 

For the particular case of HD~95086, \citet{meshkat13} used isochrone fitting to determine an age of 17$\pm$4 Myr. From our analysis, the likely age is found to be 15$\pm4$~Myr (where this age is the mean of the posterior probability distribution shown in \autoref{fpostage}). Judging by the age, this would still be a good fit to LCC, for which the average age is 16-17~Myr \citep[and the stars are known to have an age spread of around 10~Myr][]{mamajek02,preibisch08,pecaut12}. This would not be a good fit to the previous age of Carina, but it does match well with our revised analysis. In other words, the similarity between our new age for Carina and the expected age for LCC, means that HD~95086 is consistent with membership of both associations based solely on its age.

As a by-product of this process, we also infer various other stellar properties such as effective temperature and mass. Whilst these properties are not relevant to the current study, we include them in appendix \ref{sdata} as they may be of use to the community.

\subsection{Kinematic age of Carina}
\label{scarkin}

In addition to updating the age inferred from stellar evolution models, we also re-assess the kinematic age so that we can compare both using the same data and association member list. In order to re-evaluate the kinematic age of Carina, we have analysed how
the distribution of mutual distances of the member stars changes with time. We do this using the method described in \autoref{skin}, but this time tracing back the orbits of the individual member stars rather than the centre of the association to determine when they were closest together. As before, each star is cloned 1000 times with the parameters drawn from a multivariate normal distribution defined by the uncertainties on the observations.

\begin{figure}
	\centering
	\includegraphics[width=0.48\textwidth]{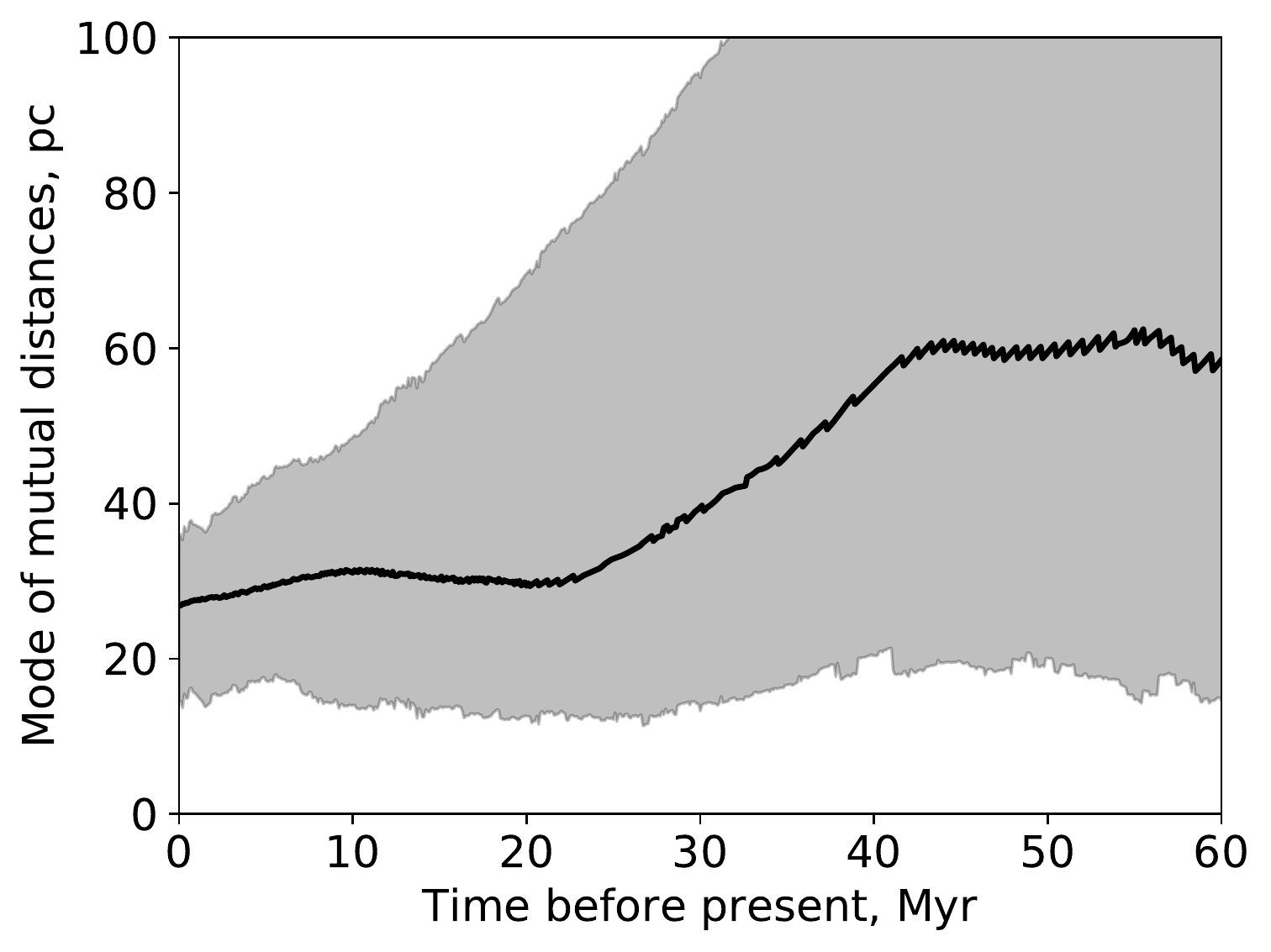}
	\caption{Mode of the distribution of the mutual distances against time for the stars that have a $>$90\% probability of being members of the Carina association. The grey region shows the 68\% highest probability density interval for each sample.} 
	\label{ftrace}
\end{figure}

The results of our trace back analysis are shown in \autoref{ftrace}. The plot shows the mode and 68\% highest probability density interval of the mutual distances between the star clones, using just the stars with a $>$90\% membership likelihood (according to our \bsig{} analysis in \autoref{scmems}). We find that as we trace back the locations of the Carina members, the average distance between the members remains roughly constant for $\sim$20~Myr before the orbits diverge, meaning that the association was born within the past 20~Myr and has exhibited little expansion since that time. We note that we also tested selecting just the members with a $>$99\% likelihood. In this case the results are broadly consistent, albeit with an increased variation in the mode. This age is somewhat smaller than the 28~Myr minimum found by \citet{miret18}, with the difference being primarily due to using a different list of members and the improvements in the \emph{Gaia} data from DR1 to DR2.

Unfortunately, we are unable to precisely determine the age through this method due to the large uncertainties, as can be seen by the grey region in \autoref{ftrace}. This is not unexpected. \citet{riedel17} simulated the expansion of an association and then used a similar method of cloning members and tracing them back to determine how accurately the age of an association can be determined through this method when assuming the accuracy of the final \emph{Gaia} data release. They found that this method could only narrow down the age to within about 10~Myr of the real age. The main issue is the uncertainty on the RVs. This is typically around 1-10\%, whilst the uncertainties on the other position and velocity parameters are around 0.01-0.1\%.

For these reasons, we take the age inferred from the stellar evolution models as the more accurate estimate of the age. That said, our kinematic age estimate does still provide extra evidence for a younger age than the previous analyses.

\section{Discussion}
\label{sdiscuss}

\subsection{The age of Carina}
In \autoref{scage}, we found that the age of Carina based on stellar evolution models is 13.3$^{+1.1}_{-0.6}$~Myr. We also found that, whilst we could not precisely determine the age from our kinematic analysis, the analysis does indicate an age younger than $\sim$20~Myr. Our two approaches give results that are, therefore, in agreement. Whilst this young age is in conflict with the work of \citet{bell15} and \citet{miret18} (who proposed ages of 45$^{+11}_{-7}$~Myr and $>$28~Myr respectively), other recent research does also point towards a younger age for Carina. For instance, \citet{schneider19} found that the lithium measurements suggest an age similar to the $\beta$ Pic Moving Group \citep[23$\pm$3~Myr according to][and references therein]{mamajek14}. \citet{ujjwal20} used a similar approach to our stellar evolution model analysis, but defined association membership using LACEwING and determined ages using MIST isochrones. Their analysis resulted in a mean age of 16.6~Myr. These results give us confidence that the Carina association is younger than previously thought.

\subsection{Planet masses}
HD~95086b was first detected by \citet{rameau13} in \emph{L'} (3.8~$\micron$) images from VLT/NaCo. Comparing the observed luminosity to `hot-start' evolutionary models \citep{baraffe03} they estimated the mass of the planet as being between 4 and 5~M$_{\rm{J}}$ based on an age estimate of 10 and 17~Myr respectively. \citet{derosa16} re-observed the planet in \emph{H} (1.5-1.8~$\micron$) and \emph{K$_1$} (1.9-2.2~$\micron$) wavebands with Gemini/GPI. They also compared their detections with the `hot-start' models of \citet{baraffe03}, finding masses of 2.7 and 4.4~\jupm{} for the \emph{H} and \emph{K$_1$} observations respectively, assuming and age of 17~Myr. In addition, they also considered `cold-start' models from \citet{mordasini13}. Comparing to these they found masses of 3.1 and $>9.1$~\jupm{} for the \emph{H} and \emph{K$_1$} observations, respectively. They did not consider younger ages but we can estimate from their Figure 10 that their \emph{H}-band photometry is consistent with a mass as low as $\sim1.5$~\jupm{} if an age of 10~Myr is considered.

The implications of a lower mass are important when considering the interaction between the planet and the debris disc. \citet{su17a} demonstrated that a planet of 4.4~\jupm{} would require an eccentricity of $\approx$0.29 if it is responsible for the inner edge of the debris disc as detected by ALMA. Alternatively a lower mass planet could exist beyond the orbit of the known planet and be responsible for clearing debris from that region of the system. In this paper we propose that this system is a member of the Carina association. The previous estimates of the age of Carina would support a more massive planet and the need for a high eccentricity or extra planet in the system could be circumvented. However, our updated age of Carina of 13.3$^{+1.1}_{-0.6}$~Myr implies a lighter mass for the planet, which reinforces the need for another planet in the system in order to explain the location of the disc's inner edge.

HD~95086 is not the only star in Carina known to host a low mass companion. A companion to HD~44627 (also known as AB Pic) was discovered by \citet{chauvin05}. The association membership of HD~44627 was originally identified as Tuc-Hor \citep{song03}. \citet{torres08} reclassified it as a member of Carina, as is supported by our results from \bsig. \citet{lee19}, however, find it to be a member of Columba. Prior to this paper, the ages of Tuc-Hor, Carina and Columba were all estimated to be $\sim$30~Myr. Based on this age, both \citet{chauvin05} and \citet{bonnefoy10} inferred a mass of $\sim$13~\jupm{} for the companion, leaving it unclear whether this companion is a planet or brown dwarf. A younger age, as proposed by our analysis, would reduce the mass required to explain the observations and clarify that this companion is most likely a gas giant rather than a brown dwarf.

\subsection{Gas in the HD~95086 debris disc}
Unlike protoplanetary discs, debris discs are gas poor. Nonetheless, in recent years detections of CO gas have been made in a number of debris disc systems \citep[see][for a recent review]{hughes18}. \citet{kral17} developed a model of CO gas released in the collisional cascade to explain the levels of gas seen in these systems. They made predictions for which discs are most likely to be good targets for detecting gas and found the HD~95086 disc to be one of the best candidates. \citet{booth19} searched this system for CO gas and found tentative evidence of its existence at an RV of 8.5$\pm$0.2~\kms{}. They demonstrated that, given the large number of channels in the data and assuming Gaussian noise, four channels would be expected to have a CO line flux equivalent to that measured. Nonetheless, the consistency between the radial profile of the gas and dust; the measured the gas mass and predicted gas mass; and the measured RV and the stellar RV found by \citet{madsen02}, all gave weight to this being a real signal. In \autoref{skin}, we demonstrated that even if it is a member of LCC, the latest data predict that the RV should then be closer to 16$\pm$4 \kms{} and in the more likely case that it is a member of Carina, the RV should be close to 17.1$\pm$0.6 \kms{}, in perfect agreement with the spectroscopic measurement of 17$\pm$2 \kms{} \citep{moor13}. From this we can conclude that the tentative detection of gas from \citet{booth19} must, in fact, just be a spurious noise signal and no gas has yet been detected in the HD~95086 debris disc.

\subsection{Carina in the context of nearby young associations}

\citet{elmegreen93} proposed a three part scenario of recent star formation close to the Sun. In this scenario, local star-forming activity began around 60~Myr ago when the Carina arm passed through the local gas. One of the associations that formed in this process resulted in the creation of the Lindblad ring, the fragmentation of which then created the nearest giant molecular clouds Orion, Sco-Cen and Perseus around 20~Myr ago. More recently a third generation formed in the region of Ophiuchus and Taurus. \citet{miret18} suggested that Tucana-Horologium, Columba and Carina likely formed as the first generation in this scenario, whilst our results show that it would have formed as part of the second generation along with the $\beta$ Pictoris moving group. Naturally, it is likely that the ages of many of the other young local associations will also be revised with new data and further analysis. For instance, \citet{goldman18} recently identified a moving group within LCC and identified four sub-groups with ages between 7-10~Myr, much younger than the 17~Myr age previously determined for LCC by \citet{pecaut16}.

\subsection{Cluster expansion}

In section \ref{skin} we considered the distance of HD~95086 from the centre of the Carina and LCC associations. This implicitly assumes that the stars in these associations formed in a gravitationally-bound cluster and this cluster is now expanding from a central point. Observations of present star formation show that stars form in a wide variety of environments from gravitationally-bound clusters to hierarchical structures and even in isolation \citep[e.g.][]{allen07}. Whilst some OB associations do show clear signs of expansion from a central point \citep{melnik20}, \citet{ward19} demonstrate that for many associations the stellar motions are not consistent with expansion from a central point, preferring hierarchical formation models and localised expansion. This means that we must be cautious in interpreting HD~95086's closer distance to the centre of Carina than LCC in the past as proof that is definitely a Carina member. Ideally this would be verified by tracing back the individual orbits of LCC members, but the large number of LCC members makes this computationally expensive and beyond the scope of this work. Nonetheless, the degree of difference between the distances to the cluster centres does give us confidence in the results of this simplified method.

\section{Conclusions}
In this paper we have analysed the age of the Carina nearby young association in light of the latest data. We began by checking the likelihood of previously proposed Carina candidates. We then applied a Bayesian stellar evolution analysis to these stars and found a mean age for the most likely candidates (membership probability $>$99\%) of 13.3$^{+1.1}_{-0.6}$~Myr. We also attempted to determine the age from a kinematic perspective by tracing back the stellar orbits. This analysis suggested that the age is likely $<$20~Myr, however, the uncertainties are too large to make any firm conclusions from the kinematics.

In the process of this work we also proposed that HD~95086 -- a star that hosts both a directly imaged planet and a resolved debris disc -- is a likely Carina member for the first time with a probability of 71\%.

Whilst we argue that HD~95086 is a member of Carina rather than LCC, the age we infer for the star is consistent with previous estimates and so the inferred mass of HD~95086b remains the same. Our reassesment of the radial velocity of the star does allow us to rule out the tentative detection of CO gas in the debris disc reported by \citet{booth19} as the radial velocity of the reported gas line is inconsistent with that of the star. 

Our finding that Carina is much younger than most previous estimates

prompts a reanalysis of the local star formation history, showing that Carina may be of a similar age to LCC.

\section*{Acknowledgements}
The authors thank Jonathan Gagn\'e and Antonio Hales for helpful discussions at an early stage of the project. The authors are grateful to the referee for constructive suggestions that helped improve the manuscript. MB and VVH acknowledge support from the Deutsche Forschungsgemeinschaft (DFG) through grants \mbox{Kr~2164/13-2} and \mbox{Ne 515/61-1} respectively. CdB acknowledges the funding of his sabbatical position through the Mexican national council for science and technology (CONACYT grant CVU No. 448248). CdB is also thankful for the support from the Jes\'us Serra Foundation Guest Program.

This research has made use of the SIMBAD database \citep{wenger00}, operated at CDS, Strasbourg, France. This work has made use of data from the European Space Agency (ESA) mission
{\it Gaia} (\url{https://www.cosmos.esa.int/gaia}), processed by the {\it Gaia}
Data Processing and Analysis Consortium (DPAC,
\url{https://www.cosmos.esa.int/web/gaia/dpac/consortium}). Funding for the DPAC
has been provided by national institutions, in particular the institutions
participating in the {\it Gaia} Multilateral Agreement. This work has made use of Topcat \citep{taylor05}.

\section*{Data availability}
The data underlying this article are available either in the article or from the Gaia Archive at \url{https://gea.esac.esa.int/archive/} using the identifiers listed in \autoref{tcarmems}. Data resulting from this work will be made available upon reasonable request.

\bibliographystyle{mnras}
\bibliography{thesis}{}

\begin{thebibliography}{}
\makeatletter
\relax
\def\mn@urlcharsother{\let\do\@makeother \do\$\do\&\do\#\do\^\do\_\do\%\do\~}
\def\mn@doi{\begingroup\mn@urlcharsother \@ifnextchar [ {\mn@doi@}
  {\mn@doi@[]}}
\def\mn@doi@[#1]#2{\def\@tempa{#1}\ifx\@tempa\@empty \href
  {http://dx.doi.org/#2} {doi:#2}\else \href {http://dx.doi.org/#2} {#1}\fi
  \endgroup}
\def\mn@eprint#1#2{\mn@eprint@#1:#2::\@nil}
\def\mn@eprint@arXiv#1{\href {http://arxiv.org/abs/#1} {{\tt arXiv:#1}}}
\def\mn@eprint@dblp#1{\href {http://dblp.uni-trier.de/rec/bibtex/#1.xml}
  {dblp:#1}}
\def\mn@eprint@#1:#2:#3:#4\@nil{\def\@tempa {#1}\def\@tempb {#2}\def\@tempc
  {#3}\ifx \@tempc \@empty \let \@tempc \@tempb \let \@tempb \@tempa \fi \ifx
  \@tempb \@empty \def\@tempb {arXiv}\fi \@ifundefined
  {mn@eprint@\@tempb}{\@tempb:\@tempc}{\expandafter \expandafter \csname
  mn@eprint@\@tempb\endcsname \expandafter{\@tempc}}}

\bibitem[\protect\citeauthoryear{{Allen} et~al.,}{{Allen}
  et~al.}{2007}]{allen07}
{Allen} L.,  et~al., 2007, in {Reipurth} B.,  {Jewitt} D.,   {Keil} K.,  eds,
  Protostars and Planets V. Univ. Arizona Press, Tucson, p.~361 (\mn@eprint
  {arXiv} {astro-ph/0603096})

\bibitem[\protect\citeauthoryear{{Ambartsumian}}{{Ambartsumian}}{1947}]{ambartsumian47}
{Ambartsumian} V.~A.,  1947, {The evolution of stars and astrophysics}.
Armenian SSR Academy of Sciences Press, Yerevan

\bibitem[\protect\citeauthoryear{{Andrae} et~al.,}{{Andrae}
  et~al.}{2018}]{andrae18}
{Andrae} R.,  et~al., 2018, \mn@doi [\aap] {10.1051/0004-6361/201732516}, \href
  {https://ui.adsabs.harvard.edu/abs/2018A&A...616A...8A} {616, A8}

\bibitem[\protect\citeauthoryear{{Bailer-Jones}, {Rybizki}, {Fouesneau},
  {Mantelet}  \& {Andrae}}{{Bailer-Jones} et~al.}{2018}]{bailer18}
{Bailer-Jones} C.~A.~L.,  {Rybizki} J.,  {Fouesneau} M.,  {Mantelet} G.,
  {Andrae} R.,  2018, \mn@doi [\aj] {10.3847/1538-3881/aacb21}, \href
  {https://ui.adsabs.harvard.edu/abs/2018AJ....156...58B} {156, 58}

\bibitem[\protect\citeauthoryear{{Bajkova} \& {Bobylev}}{{Bajkova} \&
  {Bobylev}}{2017}]{bajkova17}
{Bajkova} A.,  {Bobylev} V.,  2017, \mn@doi [Open Astronomy]
  {10.1515/astro-2017-0016}, \href
  {https://ui.adsabs.harvard.edu/abs/2017OAst...26...72B} {26, 72}

\bibitem[\protect\citeauthoryear{{Bajkova} \& {Bobylev}}{{Bajkova} \&
  {Bobylev}}{2019}]{bajkova19}
{Bajkova} A.~T.,  {Bobylev} V.~V.,  2019, \mn@doi [\mnras]
  {10.1093/mnras/stz2061}, \href
  {https://ui.adsabs.harvard.edu/abs/2019MNRAS.488.3474B} {488, 3474}

\bibitem[\protect\citeauthoryear{{Baraffe}, {Chabrier}, {Barman}, {Allard}  \&
  {Hauschildt}}{{Baraffe} et~al.}{2003}]{baraffe03}
{Baraffe} I.,  {Chabrier} G.,  {Barman} T.~S.,  {Allard} F.,   {Hauschildt}
  P.~H.,  2003, \mn@doi [\aap] {10.1051/0004-6361:20030252}, \href
  {https://ui.adsabs.harvard.edu/abs/2003A&A...402..701B} {402, 701}

\bibitem[\protect\citeauthoryear{{Bell}, {Mamajek}  \& {Naylor}}{{Bell}
  et~al.}{2015}]{bell15}
{Bell} C.~P.~M.,  {Mamajek} E.~E.,   {Naylor} T.,  2015, \mn@doi [\mnras]
  {10.1093/mnras/stv1981}, \href
  {http://adsabs.harvard.edu/abs/2015MNRAS.454..593B} {454, 593}

\bibitem[\protect\citeauthoryear{{Blaauw}}{{Blaauw}}{1964}]{blaauw64}
{Blaauw} A.,  1964, \mn@doi [\araa] {10.1146/annurev.aa.02.090164.001241},
  \href {https://ui.adsabs.harvard.edu/abs/1964ARA&A...2..213B} {2, 213}

\bibitem[\protect\citeauthoryear{{Bonnefoy}, {Chauvin}, {Rojo}, {Allard},
  {Lagrange}, {Homeier}, {Dumas}  \& {Beuzit}}{{Bonnefoy}
  et~al.}{2010}]{bonnefoy10}
{Bonnefoy} M.,  {Chauvin} G.,  {Rojo} P.,  {Allard} F.,  {Lagrange} A.~M.,
  {Homeier} D.,  {Dumas} C.,   {Beuzit} J.~L.,  2010, \mn@doi [\aap]
  {10.1051/0004-6361/200912688}, \href
  {https://ui.adsabs.harvard.edu/abs/2010A&A...512A..52B} {512, A52}

\bibitem[\protect\citeauthoryear{{Booth} et~al.,}{{Booth}
  et~al.}{2019}]{booth19}
{Booth} M.,  et~al., 2019, \mn@doi [\mnras] {10.1093/mnras/sty2993}, \href
  {http://adsabs.harvard.edu/abs/2019MNRAS.482.3443B} {482, 3443}

\bibitem[\protect\citeauthoryear{{Bressan}, {Marigo}, {Girardi}, {Salasnich},
  {Dal Cero}, {Rubele}  \& {Nanni}}{{Bressan} et~al.}{2012}]{bressan12}
{Bressan} A.,  {Marigo} P.,  {Girardi} L.,  {Salasnich} B.,  {Dal Cero} C.,
  {Rubele} S.,   {Nanni} A.,  2012, \mn@doi [\mnras]
  {10.1111/j.1365-2966.2012.21948.x}, \href
  {https://ui.adsabs.harvard.edu/abs/2012MNRAS.427..127B} {427, 127}

\bibitem[\protect\citeauthoryear{{Chauvin} et~al.,}{{Chauvin}
  et~al.}{2005}]{chauvin05}
{Chauvin} G.,  et~al., 2005, \mn@doi [\aap] {10.1051/0004-6361:200500111},
  \href {https://ui.adsabs.harvard.edu/abs/2005A&A...438L..29C} {438, L29}

\bibitem[\protect\citeauthoryear{{Chen}, {Pecaut}, {Mamajek}, {Su}  \&
  {Bitner}}{{Chen} et~al.}{2012}]{chen12}
{Chen} C.~H.,  {Pecaut} M.,  {Mamajek} E.~E.,  {Su} K.~Y.~L.,   {Bitner} M.,
  2012, \mn@doi [\apj] {10.1088/0004-637X/756/2/133}, \href
  {http://adsabs.harvard.edu/abs/2012ApJ...756..133C} {756, 133}

\bibitem[\protect\citeauthoryear{{Chen}, {Girardi}, {Bressan}, {Marigo},
  {Barbieri}  \& {Kong}}{{Chen} et~al.}{2014}]{chen14}
{Chen} Y.,  {Girardi} L.,  {Bressan} A.,  {Marigo} P.,  {Barbieri} M.,   {Kong}
  X.,  2014, \mn@doi [\mnras] {10.1093/mnras/stu1605}, \href
  {https://ui.adsabs.harvard.edu/abs/2014MNRAS.444.2525C} {444, 2525}

\bibitem[\protect\citeauthoryear{{Chen}, {Bressan}, {Girardi}, {Marigo}, {Kong}
   \& {Lanza}}{{Chen} et~al.}{2015}]{chen15}
{Chen} Y.,  {Bressan} A.,  {Girardi} L.,  {Marigo} P.,  {Kong} X.,   {Lanza}
  A.,  2015, \mn@doi [\mnras] {10.1093/mnras/stv1281}, \href
  {https://ui.adsabs.harvard.edu/abs/2015MNRAS.452.1068C} {452, 1068}

\bibitem[\protect\citeauthoryear{{Damiani}, {Prisinzano}, {Pillitteri},
  {Micela}  \& {Sciortino}}{{Damiani} et~al.}{2019}]{damiani19}
{Damiani} F.,  {Prisinzano} L.,  {Pillitteri} I.,  {Micela} G.,   {Sciortino}
  S.,  2019, \mn@doi [\aap] {10.1051/0004-6361/201833994}, \href
  {https://ui.adsabs.harvard.edu/abs/2019A&A...623A.112D} {623, A112}

\bibitem[\protect\citeauthoryear{{De Rosa} et~al.,}{{De Rosa}
  et~al.}{2016}]{derosa16}
{De Rosa} R.~J.,  et~al., 2016, \mn@doi [\apj] {10.3847/0004-637X/824/2/121},
  \href {https://ui.adsabs.harvard.edu/abs/2016ApJ...824..121D} {824, 121}

\bibitem[\protect\citeauthoryear{{Durkan} et~al.,}{{Durkan}
  et~al.}{2018}]{durkan18}
{Durkan} S.,  et~al., 2018, \mn@doi [\aap] {10.1051/0004-6361/201732156}, \href
  {https://ui.adsabs.harvard.edu/abs/2018A&A...618A...5D} {618, A5}

\bibitem[\protect\citeauthoryear{{Eker} et~al.,}{{Eker} et~al.}{2018}]{eker18}
{Eker} Z.,  et~al., 2018, \mn@doi [\mnras] {10.1093/mnras/sty1834}, \href
  {https://ui.adsabs.harvard.edu/abs/2018MNRAS.479.5491E} {479, 5491}

\bibitem[\protect\citeauthoryear{{Elliott} et~al.,}{{Elliott}
  et~al.}{2015}]{elliott15}
{Elliott} P.,  et~al., 2015, \mn@doi [\aap] {10.1051/0004-6361/201525794},
  \href {https://ui.adsabs.harvard.edu/abs/2015A&A...580A..88E} {580, A88}

\bibitem[\protect\citeauthoryear{{Elmegreen}}{{Elmegreen}}{1993}]{elmegreen93}
{Elmegreen} B.~G.,  1993, in {Levy} E.~H.,  {Lunine} J.~I.,  eds, Protostars
  and Planets III. p.~97

\bibitem[\protect\citeauthoryear{{Fern{\'a}ndez}, {Figueras}  \&
  {Torra}}{{Fern{\'a}ndez} et~al.}{2008}]{fernandez08}
{Fern{\'a}ndez} D.,  {Figueras} F.,   {Torra} J.,  2008, \mn@doi [\aap]
  {10.1051/0004-6361:20077720}, \href
  {https://ui.adsabs.harvard.edu/abs/2008A&A...480..735F} {480, 735}

\bibitem[\protect\citeauthoryear{{Gagn{\'e}} \& {Faherty}}{{Gagn{\'e}} \&
  {Faherty}}{2018}]{gagne18b}
{Gagn{\'e}} J.,  {Faherty} J.~K.,  2018, \mn@doi [\apj]
  {10.3847/1538-4357/aaca2e}, \href
  {https://ui.adsabs.harvard.edu/abs/2018ApJ...862..138G} {862, 138}

\bibitem[\protect\citeauthoryear{{Gagn{\'e}}, {Lafreni{\`e}re}, {Doyon}, {Malo}
   \& {Artigau}}{{Gagn{\'e}} et~al.}{2014}]{gagne14}
{Gagn{\'e}} J.,  {Lafreni{\`e}re} D.,  {Doyon} R.,  {Malo} L.,   {Artigau}
  {\'E}.,  2014, \mn@doi [\apj] {10.1088/0004-637X/783/2/121}, \href
  {https://ui.adsabs.harvard.edu/abs/2014ApJ...783..121G} {783, 121}

\bibitem[\protect\citeauthoryear{{Gagn{\'e}} et~al.,}{{Gagn{\'e}}
  et~al.}{2018a}]{gagne18}
{Gagn{\'e}} J.,  et~al., 2018a, \mn@doi [\apj] {10.3847/1538-4357/aaae09},
  \href {https://ui.adsabs.harvard.edu/\#abs/2018ApJ...856...23G} {856, 23}

\bibitem[\protect\citeauthoryear{{Gagn{\'e}}, {Roy-Loubier}, {Faherty}, {Doyon}
   \& {Malo}}{{Gagn{\'e}} et~al.}{2018b}]{gagne18a}
{Gagn{\'e}} J.,  {Roy-Loubier} O.,  {Faherty} J.~K.,  {Doyon} R.,   {Malo} L.,
  2018b, \mn@doi [\apj] {10.3847/1538-4357/aac2b8}, \href
  {https://ui.adsabs.harvard.edu/abs/2018ApJ...860...43G} {860, 43}

\bibitem[\protect\citeauthoryear{{Gaia Collaboration} et~al.,}{{Gaia
  Collaboration} et~al.}{2018}]{gaia18}
{Gaia Collaboration} et~al., 2018, \mn@doi [\aap]
  {10.1051/0004-6361/201833051}, \href
  {http://adsabs.harvard.edu/abs/2018A%26A...616A...1G} {616, A1}

\bibitem[\protect\citeauthoryear{{Goldman}, {R{\"o}ser}, {Schilbach},
  {Mo{\'o}r}  \& {Henning}}{{Goldman} et~al.}{2018}]{goldman18}
{Goldman} B.,  {R{\"o}ser} S.,  {Schilbach} E.,  {Mo{\'o}r} A.~C.,   {Henning}
  T.,  2018, \mn@doi [\apj] {10.3847/1538-4357/aae64c}, \href
  {https://ui.adsabs.harvard.edu/abs/2018ApJ...868...32G} {868, 32}

\bibitem[\protect\citeauthoryear{{Gontcharov}}{{Gontcharov}}{2006}]{gontcharov06}
{Gontcharov} G.~A.,  2006, \mn@doi [Astronomy Letters]
  {10.1134/S1063773706110065}, \href
  {http://adsabs.harvard.edu/abs/2006AstL...32..759G} {32, 759}

\bibitem[\protect\citeauthoryear{{Gray}, {Corbally}, {Garrison}, {McFadden},
  {Bubar}, {McGahee}, {O'Donoghue}  \& {Knox}}{{Gray} et~al.}{2006}]{gray06}
{Gray} R.~O.,  {Corbally} C.~J.,  {Garrison} R.~F.,  {McFadden} M.~T.,  {Bubar}
  E.~J.,  {McGahee} C.~E.,  {O'Donoghue} A.~A.,   {Knox} E.~R.,  2006, \mn@doi
  [\aj] {10.1086/504637}, \href
  {http://cdsads.u-strasbg.fr/abs/2006AJ....132..161G} {132, 161}

\bibitem[\protect\citeauthoryear{{Houk} \& {Cowley}}{{Houk} \&
  {Cowley}}{1975}]{houk75}
{Houk} N.,  {Cowley} A.~P.,  1975, {University of Michigan Catalogue of
  two-dimensional spectral types for the HD stars. Volume I.}.
Dept. of Astronomy, University of Michigan

\bibitem[\protect\citeauthoryear{{Hughes}, {Duch{\^e}ne}  \&
  {Matthews}}{{Hughes} et~al.}{2018}]{hughes18}
{Hughes} A.~M.,  {Duch{\^e}ne} G.,   {Matthews} B.~C.,  2018, \mn@doi [\araa]
  {10.1146/annurev-astro-081817-052035}, \href
  {http://adsabs.harvard.edu/abs/2018ARA%26A..56..541H} {56, 541}

\bibitem[\protect\citeauthoryear{{Kharchenko}, {Scholz}, {Piskunov},
  {R{\"o}ser}  \& {Schilbach}}{{Kharchenko} et~al.}{2007}]{kharchenko07}
{Kharchenko} N.~V.,  {Scholz} R.-D.,  {Piskunov} A.~E.,  {R{\"o}ser} S.,
  {Schilbach} E.,  2007, \mn@doi [Astronomische Nachrichten]
  {10.1002/asna.200710776}, \href
  {http://adsabs.harvard.edu/abs/2007AN....328..889K} {328, 889}

\bibitem[\protect\citeauthoryear{{Kral}, {Matr{\`a}}, {Wyatt}  \&
  {Kennedy}}{{Kral} et~al.}{2017}]{kral17}
{Kral} Q.,  {Matr{\`a}} L.,  {Wyatt} M.~C.,   {Kennedy} G.~M.,  2017, \mn@doi
  [\mnras] {10.1093/mnras/stx730}, \href
  {http://adsabs.harvard.edu/abs/2017MNRAS.469..521K} {469, 521}

\bibitem[\protect\citeauthoryear{{Krivov} \& {Booth}}{{Krivov} \&
  {Booth}}{2018}]{krivov18b}
{Krivov} A.~V.,  {Booth} M.,  2018, \mn@doi [\mnras] {10.1093/mnras/sty1607},
  \href {http://adsabs.harvard.edu/abs/2018MNRAS.479.3300K} {479, 3300}

\bibitem[\protect\citeauthoryear{{Krivov}, {Ide}, {L{\"o}hne}, {Johansen}  \&
  {Blum}}{{Krivov} et~al.}{2018}]{krivov18a}
{Krivov} A.~V.,  {Ide} A.,  {L{\"o}hne} T.,  {Johansen} A.,   {Blum} J.,  2018,
  \mn@doi [\mnras] {10.1093/mnras/stx2932}, \href
  {http://adsabs.harvard.edu/abs/2018MNRAS.474.2564K} {474, 2564}

\bibitem[\protect\citeauthoryear{{Lee} \& {Song}}{{Lee} \&
  {Song}}{2019}]{lee19}
{Lee} J.,  {Song} I.,  2019, \mn@doi [\mnras] {10.1093/mnras/stz1044}, \href
  {https://ui.adsabs.harvard.edu/abs/2019MNRAS.486.3434L} {486, 3434}

\bibitem[\protect\citeauthoryear{{Madsen}, {Dravins}  \& {Lindegren}}{{Madsen}
  et~al.}{2002}]{madsen02}
{Madsen} S.,  {Dravins} D.,   {Lindegren} L.,  2002, \mn@doi [\aap]
  {10.1051/0004-6361:20011458}, \href
  {http://adsabs.harvard.edu/abs/2002A%26A...381..446M} {381, 446}

\bibitem[\protect\citeauthoryear{{Ma{\'\i}z Apell{\'a}niz} \&
  {Weiler}}{{Ma{\'\i}z Apell{\'a}niz} \& {Weiler}}{2018}]{maiz18}
{Ma{\'\i}z Apell{\'a}niz} J.,  {Weiler} M.,  2018, \mn@doi [\aap]
  {10.1051/0004-6361/201834051}, \href
  {https://ui.adsabs.harvard.edu/abs/2018A&A...619A.180M} {619, A180}

\bibitem[\protect\citeauthoryear{{Malo}, {Doyon}, {Lafreni{\`e}re}, {Artigau},
  {Gagn{\'e}}, {Baron}  \& {Riedel}}{{Malo} et~al.}{2013}]{malo13}
{Malo} L.,  {Doyon} R.,  {Lafreni{\`e}re} D.,  {Artigau} {\'E}.,  {Gagn{\'e}}
  J.,  {Baron} F.,   {Riedel} A.,  2013, \mn@doi [\apj]
  {10.1088/0004-637X/762/2/88}, \href
  {https://ui.adsabs.harvard.edu/abs/2013ApJ...762...88M} {762, 88}

\bibitem[\protect\citeauthoryear{{Malo}, {Artigau}, {Doyon}, {Lafreni{\`e}re},
  {Albert}  \& {Gagn{\'e}}}{{Malo} et~al.}{2014}]{malo14}
{Malo} L.,  {Artigau} {\'E}.,  {Doyon} R.,  {Lafreni{\`e}re} D.,  {Albert} L.,
   {Gagn{\'e}} J.,  2014, \mn@doi [\apj] {10.1088/0004-637X/788/1/81}, \href
  {https://ui.adsabs.harvard.edu/abs/2014ApJ...788...81M} {788, 81}

\bibitem[\protect\citeauthoryear{{Mamajek} \& {Bell}}{{Mamajek} \&
  {Bell}}{2014}]{mamajek14}
{Mamajek} E.~E.,  {Bell} C. P.~M.,  2014, \mn@doi [\mnras]
  {10.1093/mnras/stu1894}, \href
  {https://ui.adsabs.harvard.edu/abs/2014MNRAS.445.2169M} {445, 2169}

\bibitem[\protect\citeauthoryear{{Mamajek}, {Meyer}  \& {Liebert}}{{Mamajek}
  et~al.}{2002}]{mamajek02}
{Mamajek} E.~E.,  {Meyer} M.~R.,   {Liebert} J.,  2002, \mn@doi [\aj]
  {10.1086/341952}, \href
  {https://ui.adsabs.harvard.edu/abs/2002AJ....124.1670M} {124, 1670}

\bibitem[\protect\citeauthoryear{{Mason}, {Wycoff}, {Hartkopf}, {Douglass}  \&
  {Worley}}{{Mason} et~al.}{2001}]{mason01}
{Mason} B.~D.,  {Wycoff} G.~L.,  {Hartkopf} W.~I.,  {Douglass} G.~G.,
  {Worley} C.~E.,  2001, \mn@doi [\aj] {10.1086/323920}, \href
  {https://ui.adsabs.harvard.edu/abs/2001AJ....122.3466M} {122, 3466}

\bibitem[\protect\citeauthoryear{{Melnik} \& {Dambis}}{{Melnik} \&
  {Dambis}}{2020}]{melnik20}
{Melnik} A.~M.,  {Dambis} A.~K.,  2020, \mn@doi [\mnras]
  {10.1093/mnras/staa454}, \href
  {https://ui.adsabs.harvard.edu/abs/2020MNRAS.493.2339M} {493, 2339}

\bibitem[\protect\citeauthoryear{{Meshkat} et~al.,}{{Meshkat}
  et~al.}{2013}]{meshkat13}
{Meshkat} T.,  et~al., 2013, \mn@doi [\apjl] {10.1088/2041-8205/775/2/L40},
  \href {http://adsabs.harvard.edu/abs/2013ApJ...775L..40M} {775, L40}

\bibitem[\protect\citeauthoryear{{Miret-Roig}, {Antoja}, {Romero-G{\'o}mez}  \&
  {Figueras}}{{Miret-Roig} et~al.}{2018}]{miret18}
{Miret-Roig} N.,  {Antoja} T.,  {Romero-G{\'o}mez} M.,   {Figueras} F.,  2018,
  \mn@doi [\aap] {10.1051/0004-6361/201731976}, \href
  {https://ui.adsabs.harvard.edu/abs/2018A&A...615A..51M} {615, A51}

\bibitem[\protect\citeauthoryear{{Mo{\'o}r}, {Szab{\'o}}, {Kiss}, {Kiss},
  {{\'A}brah{\'a}m}, {Szul{\'a}gyi}, {K{\'o}sp{\'a}l}  \& {Szalai}}{{Mo{\'o}r}
  et~al.}{2013a}]{moor13b}
{Mo{\'o}r} A.,  {Szab{\'o}} G.~M.,  {Kiss} L.~L.,  {Kiss} C.,
  {{\'A}brah{\'a}m} P.,  {Szul{\'a}gyi} J.,  {K{\'o}sp{\'a}l} {\'A}.,
  {Szalai} T.,  2013a, \mn@doi [\mnras] {10.1093/mnras/stt1381}, \href
  {https://ui.adsabs.harvard.edu/abs/2013MNRAS.435.1376M} {435, 1376}

\bibitem[\protect\citeauthoryear{{Mo{\'o}r} et~al.,}{{Mo{\'o}r}
  et~al.}{2013b}]{moor13}
{Mo{\'o}r} A.,  et~al., 2013b, \mn@doi [\apjl] {10.1088/2041-8205/775/2/L51},
  \href {http://adsabs.harvard.edu/abs/2013ApJ...775L..51M} {775, L51}

\bibitem[\protect\citeauthoryear{{Mordasini}}{{Mordasini}}{2013}]{mordasini13}
{Mordasini} C.,  2013, \mn@doi [\aap] {10.1051/0004-6361/201321617}, \href
  {https://ui.adsabs.harvard.edu/abs/2013A&A...558A.113M} {558, A113}

\bibitem[\protect\citeauthoryear{{Murphy}, {Lawson}  \& {Bento}}{{Murphy}
  et~al.}{2015}]{murphy15}
{Murphy} S.~J.,  {Lawson} W.~A.,   {Bento} J.,  2015, \mn@doi [\mnras]
  {10.1093/mnras/stv1745}, \href
  {https://ui.adsabs.harvard.edu/abs/2015MNRAS.453.2220M} {453, 2220}

\bibitem[\protect\citeauthoryear{{Murphy}, {Mamajek}  \& {Bell}}{{Murphy}
  et~al.}{2018}]{murphy18}
{Murphy} S.~J.,  {Mamajek} E.~E.,   {Bell} C. P.~M.,  2018, \mn@doi [\mnras]
  {10.1093/mnras/sty471}, \href
  {https://ui.adsabs.harvard.edu/abs/2018MNRAS.476.3290M} {476, 3290}

\bibitem[\protect\citeauthoryear{{Neuh{\"a}user}, {Gie{\ss}ler}  \&
  {Hambaryan}}{{Neuh{\"a}user} et~al.}{2019}]{neuhauser19}
{Neuh{\"a}user} R.,  {Gie{\ss}ler} F.,   {Hambaryan} V.~V.,  2019, \mn@doi
  [\mnras] {10.1093/mnras/stz2629}, \href
  {https://ui.adsabs.harvard.edu/abs/2019MNRAS.tmp.2261N} {p.~2261}

\bibitem[\protect\citeauthoryear{{Palous}, {Jungwiert}  \& {Kopecky}}{{Palous}
  et~al.}{1993}]{palous93}
{Palous} J.,  {Jungwiert} B.,   {Kopecky} J.,  1993, \aap, \href
  {https://ui.adsabs.harvard.edu/abs/1993A&A...274..189P} {274, 189}

\bibitem[\protect\citeauthoryear{{Pecaut} \& {Mamajek}}{{Pecaut} \&
  {Mamajek}}{2013}]{pecaut13}
{Pecaut} M.~J.,  {Mamajek} E.~E.,  2013, \mn@doi [\apjs]
  {10.1088/0067-0049/208/1/9}, \href
  {https://ui.adsabs.harvard.edu/abs/2013ApJS..208....9P} {208, 9}

\bibitem[\protect\citeauthoryear{{Pecaut} \& {Mamajek}}{{Pecaut} \&
  {Mamajek}}{2016}]{pecaut16}
{Pecaut} M.~J.,  {Mamajek} E.~E.,  2016, \mn@doi [\mnras]
  {10.1093/mnras/stw1300}, \href
  {https://ui.adsabs.harvard.edu/abs/2016MNRAS.461..794P} {461, 794}

\bibitem[\protect\citeauthoryear{{Pecaut}, {Mamajek}  \& {Bubar}}{{Pecaut}
  et~al.}{2012}]{pecaut12}
{Pecaut} M.~J.,  {Mamajek} E.~E.,   {Bubar} E.~J.,  2012, \mn@doi [\apj]
  {10.1088/0004-637X/746/2/154}, \href
  {https://ui.adsabs.harvard.edu/abs/2012ApJ...746..154P} {746, 154}

\bibitem[\protect\citeauthoryear{{Preibisch} \& {Mamajek}}{{Preibisch} \&
  {Mamajek}}{2008}]{preibisch08}
{Preibisch} T.,  {Mamajek} E.,  2008, Handbook of Star Forming Regions: Volume
  II, The Southern Sky.
Astronomical Society of the Pacific, p.~235

\bibitem[\protect\citeauthoryear{{Rameau} et~al.,}{{Rameau}
  et~al.}{2013}]{rameau13}
{Rameau} J.,  et~al., 2013, \mn@doi [\apjl] {10.1088/2041-8205/772/2/L15},
  \href {http://adsabs.harvard.edu/abs/2013ApJ...772L..15R} {772, L15}

\bibitem[\protect\citeauthoryear{{Riedel}, {Blunt}, {Lambrides}, {Rice}, {Cruz}
   \& {Faherty}}{{Riedel} et~al.}{2017}]{riedel17}
{Riedel} A.~R.,  {Blunt} S.~C.,  {Lambrides} E.~L.,  {Rice} E.~L.,  {Cruz}
  K.~L.,   {Faherty} J.~K.,  2017, \mn@doi [\aj] {10.3847/1538-3881/153/3/95},
  \href {https://ui.adsabs.harvard.edu/abs/2017AJ....153...95R} {153, 95}

\bibitem[\protect\citeauthoryear{{Rizzuto}, {Ireland}  \& {Zucker}}{{Rizzuto}
  et~al.}{2012}]{rizzuto12}
{Rizzuto} A.~C.,  {Ireland} M.~J.,   {Zucker} D.~B.,  2012, \mn@doi [\mnras]
  {10.1111/j.1745-3933.2012.01214.x}, \href
  {http://adsabs.harvard.edu/abs/2012MNRAS.421L..97R} {421, L97}

\bibitem[\protect\citeauthoryear{{Sartoretti} et~al.,}{{Sartoretti}
  et~al.}{2018}]{sartoretti18}
{Sartoretti} P.,  et~al., 2018, \mn@doi [\aap] {10.1051/0004-6361/201832836},
  \href {https://ui.adsabs.harvard.edu/abs/2018A&A...616A...6S} {616, A6}

\bibitem[\protect\citeauthoryear{{Schneider}, {Shkolnik}, {Allers}, {Kraus},
  {Liu}, {Weinberger}  \& {Flagg}}{{Schneider} et~al.}{2019}]{schneider19}
{Schneider} A.~C.,  {Shkolnik} E.~L.,  {Allers} K.~N.,  {Kraus} A.~L.,  {Liu}
  M.~C.,  {Weinberger} A.~J.,   {Flagg} L.,  2019, \mn@doi [\aj]
  {10.3847/1538-3881/ab1a26}, \href
  {https://ui.adsabs.harvard.edu/abs/2019AJ....157..234S} {157, 234}

\bibitem[\protect\citeauthoryear{{Silverberg} et~al.,}{{Silverberg}
  et~al.}{2016}]{silverberg16}
{Silverberg} S.~M.,  et~al., 2016, \mn@doi [\apjl]
  {10.3847/2041-8205/830/2/L28}, \href
  {https://ui.adsabs.harvard.edu/abs/2016ApJ...830L..28S} {830, L28}

\bibitem[\protect\citeauthoryear{{Song}, {Zuckerman}  \& {Bessell}}{{Song}
  et~al.}{2003}]{song03}
{Song} I.,  {Zuckerman} B.,   {Bessell} M.~S.,  2003, \mn@doi [\apj]
  {10.1086/379194}, \href
  {https://ui.adsabs.harvard.edu/abs/2003ApJ...599..342S} {599, 342}

\bibitem[\protect\citeauthoryear{{Su} et~al.,}{{Su} et~al.}{2017}]{su17a}
{Su} K.~Y.~L.,  et~al., 2017, \mn@doi [\aj] {10.3847/1538-3881/aa906b}, \href
  {http://adsabs.harvard.edu/abs/2017AJ....154..225S} {154, 225}

\bibitem[\protect\citeauthoryear{{Tang}, {Bressan}, {Rosenfield}, {Slemer},
  {Marigo}, {Girardi}  \& {Bianchi}}{{Tang} et~al.}{2014}]{tang14}
{Tang} J.,  {Bressan} A.,  {Rosenfield} P.,  {Slemer} A.,  {Marigo} P.,
  {Girardi} L.,   {Bianchi} L.,  2014, \mn@doi [\mnras]
  {10.1093/mnras/stu2029}, \href
  {https://ui.adsabs.harvard.edu/abs/2014MNRAS.445.4287T} {445, 4287}

\bibitem[\protect\citeauthoryear{{Taylor}}{{Taylor}}{2005}]{taylor05}
{Taylor} M.~B.,  2005, in {Shopbell} P.,  {Britton} M.,   {Ebert} R.,  eds,
  Astronomical Society of the Pacific Conference Series Vol. 347, Astronomical
  Data Analysis Software and Systems XIV. p.~29

\bibitem[\protect\citeauthoryear{{Torres}, {Quast}, {de La Reza}, {da Silva}
  \& {Melo}}{{Torres} et~al.}{2001}]{torres01}
{Torres} C.~A.~O.,  {Quast} G.~R.,  {de La Reza} R.,  {da Silva} L.,   {Melo}
  C.~H.~F.,  2001, in {Jayawardhana} R.,  {Greene} T.,  eds,  Astronomical
  Society of the Pacific Conference Series Vol. 244, Young Stars Near Earth:
  Progress and Prospects. p.~43 (\mn@eprint {arXiv} {astro-ph/0105291})

\bibitem[\protect\citeauthoryear{{Torres}, {Quast}, {da Silva}, {de La Reza},
  {Melo}  \& {Sterzik}}{{Torres} et~al.}{2006}]{torres06}
{Torres} C.~A.~O.,  {Quast} G.~R.,  {da Silva} L.,  {de La Reza} R.,  {Melo}
  C.~H.~F.,   {Sterzik} M.,  2006, \mn@doi [\aap] {10.1051/0004-6361:20065602},
  \href {http://cdsads.u-strasbg.fr/abs/2006A%26A...460..695T} {460, 695}

\bibitem[\protect\citeauthoryear{{Torres}, {Quast}, {Melo}  \&
  {Sterzik}}{{Torres} et~al.}{2008}]{torres08}
{Torres} C.~A.~O.,  {Quast} G.~R.,  {Melo} C.~H.~F.,   {Sterzik} M.~F.,  2008,
  Handbook of Star Forming Regions: Volume II, The Southern Sky.
Astronomical Society of the Pacific, p.~757

\bibitem[\protect\citeauthoryear{{Ujjwal}, {Kartha}, {Mathew}, {Manoj}  \&
  {Narang}}{{Ujjwal} et~al.}{2020}]{ujjwal20}
{Ujjwal} K.,  {Kartha} S.~S.,  {Mathew} B.,  {Manoj} P.,   {Narang} M.,  2020,
  \mn@doi [\aj] {10.3847/1538-3881/ab76d6}, \href
  {https://ui.adsabs.harvard.edu/abs/2020AJ....159..166U} {159, 166}

\bibitem[\protect\citeauthoryear{{Vican}}{{Vican}}{2012}]{vican12}
{Vican} L.,  2012, \mn@doi [\aj] {10.1088/0004-6256/143/6/135}, \href
  {http://adsabs.harvard.edu/abs/2012AJ....143..135V} {143, 135}

\bibitem[\protect\citeauthoryear{{Ward}, {Kruijssen}  \& {Rix}}{{Ward}
  et~al.}{2020}]{ward19}
{Ward} J.~L.,  {Kruijssen} J.~M.~D.,   {Rix} H.-W.,  2020, \mn@doi [\mnras]
  {10.1093/mnras/staa1056}, \href
  {https://ui.adsabs.harvard.edu/abs/2020MNRAS.495..663W} {495, 663}

\bibitem[\protect\citeauthoryear{{Wenger} et~al.,}{{Wenger}
  et~al.}{2000}]{wenger00}
{Wenger} M.,  et~al., 2000, \mn@doi [\aaps] {10.1051/aas:2000332}, \href
  {https://ui.adsabs.harvard.edu/abs/2000A&AS..143....9W} {143, 9}

\bibitem[\protect\citeauthoryear{{Zuckerman} \& {Song}}{{Zuckerman} \&
  {Song}}{2004}]{zuckerman04a}
{Zuckerman} B.,  {Song} I.,  2004, \mn@doi [\araa]
  {10.1146/annurev.astro.42.053102.134111}, \href
  {https://ui.adsabs.harvard.edu/abs/2004ARA&A..42..685Z} {42, 685}

\bibitem[\protect\citeauthoryear{{de Zeeuw}, {Hoogerwerf}, {de Bruijne},
  {Brown}  \& {Blaauw}}{{de Zeeuw} et~al.}{1999}]{dezeeuw99}
{de Zeeuw} P.~T.,  {Hoogerwerf} R.,  {de Bruijne} J.~H.~J.,  {Brown} A.~G.~A.,
   {Blaauw} A.,  1999, \mn@doi [\aj] {10.1086/300682}, \href
  {http://adsabs.harvard.edu/abs/1999AJ....117..354D} {117, 354}

\bibitem[\protect\citeauthoryear{{del Burgo} \& {Allende Prieto}}{{del Burgo}
  \& {Allende Prieto}}{2018}]{delburgo18}
{del Burgo} C.,  {Allende Prieto} C.,  2018, \mn@doi [\mnras]
  {10.1093/mnras/sty1371}, \href
  {https://ui.adsabs.harvard.edu/abs/2018MNRAS.479.1953D} {479, 1953}

\makeatother
\end{thebibliography}

\appendix

\section{Radial velocities from Kharchenko et al. (2007)}
\label{skhar}
In the process of this work we noticed that there is an issue with the catalogue of \citet{kharchenko07}. \citet{kharchenko07} compiled RVs from a number of other sources in the literature including \citet{gontcharov06}. However, the source of many of the stars is incorrect. There are 396 stars in \citet{kharchenko07} that are listed as coming from the catalogue of \citet{gontcharov06} but do not actually appear in that catalogue. They likely came from other literature sources. Most intriguingly, 350 of them have RVs that match those in \citet{madsen02}. Unlike \citet{gontcharov06}, \citet{madsen02} did not determine RVs by observing stars spectroscopically. Instead, they made use of membership lists of young associations and determined, based on the Hipparcos parallaxes and proper motions, what RVs stars should have if they are in the given young associations. Confusing RVs measured in this way with those measured by spectroscopy is problematic as firstly the assumption of membership in a young association may not be correct and, secondly, even if it is, the true RV of a star may be somewhat different from that expected due to interactions with other stars. This has previously be noted by \citet{murphy15}, although they did not specifically note that HD~95086 is one of the stars affected.

\section{Stellar data}
\label{sdata}
In addition to the ages, the code of \citet{delburgo18} also infers various other properties of the stars. Whilst not relevant to current study, we include these in \autoref{tsprops} as they may be useful to the community for other purposes.

\begin{table*}
\begin{tabular}{lcccc}
                     Name &     log$L$ &      $T_{\rm{eff}}$, K &     $M_{\star}$, M$_{\odot}$    &  $R_{\star}$, R$_{\odot}$ \\ 
\hline
            HD 49855	&	-0.224	$\pm$	0.006	&	5430	$\pm$	30	&	0.95	$\pm$	0.05	&	0.873	$\pm$	0.015	\\
      UCAC3 53-40215	&	-0.854	$\pm$	0.005	&	3799	$\pm$	21	&	0.69	$\pm$	0.03	&	0.864	$\pm$	0.012	\\
2MASS J08040534-6316396	&	-1.015	$\pm$	0.006	&	3470	$\pm$	50	&	0.58	$\pm$	0.09	&	0.860	$\pm$	0.022	\\
2MASS J09315840-6209258	&	-1.008	$\pm$	0.013	&	3260	$\pm$	50	&	0.45	$\pm$	0.09	&	0.985	$\pm$	0.018	\\
            HD 42270	&	-0.131	$\pm$	0.008	&	5280	$\pm$	70	&	1.003	$\pm$	0.014	&	1.03	$\pm$	0.03	\\
2MASS J08063608-7444249	&	-1.140	$\pm$	0.004	&	3500	$\pm$	30	&	0.57	$\pm$	0.05	&	0.732	$\pm$	0.013	\\
            HD 37402	&	0.246	$\pm$	0.010	&	6231	$\pm$	11	&	1.21	$\pm$	0.04	&	1.139	$\pm$	0.013	\\
           HD 298936	&	-0.384	$\pm$	0.009	&	4660	$\pm$	50	&	0.923	$\pm$	0.029	&	0.99	$\pm$	0.03	\\
               m Car	&	1.877	$\pm$	0.017	&	10470	$\pm$	120	&	2.84	$\pm$	0.22	&	2.64	$\pm$	0.04	\\
2MASS J06262199-7516404	&	-1.081	$\pm$	0.004	&	3690	$\pm$	50	&	0.61	$\pm$	0.03	&	0.705	$\pm$	0.018	\\
         V* V479 Car	&	-0.020	$\pm$	0.007	&	5130	$\pm$	50	&	1.151	$\pm$	0.016	&	1.24	$\pm$	0.03	\\
            HD 44627	&	-0.227	$\pm$	0.006	&	5020	$\pm$	40	&	0.974	$\pm$	0.014	&	1.020	$\pm$	0.023	\\
            HD 55279	&	-0.372	$\pm$	0.006	&	4800	$\pm$	40	&	0.903	$\pm$	0.013	&	0.942	$\pm$	0.022	\\
              AL 442	&	-1.121	$\pm$	0.024	&	3140	$\pm$	30	&	0.36	$\pm$	0.07	&	0.933	$\pm$	0.023	\\
2MASS J07441105-6458052	&	-1.094	$\pm$	0.005	&	3915	$\pm$	22	&	0.636	$\pm$	0.016	&	0.617	$\pm$	0.006	\\
2MASS J09180165-5452332	&	-1.23	$\pm$	0.04	&	2938	$\pm$	13	&	0.23	$\pm$	0.06	&	0.94	$\pm$	0.04	\\
2MASS J07065772-5353463	&	-0.836	$\pm$	0.004	&	3816	$\pm$	27	&	0.68	$\pm$	0.04	&	0.874	$\pm$	0.014	\\
WISE J080822.18-644357.3  	&	-2.151	$\pm$	0.011	&	3172	$\pm$	16	&	0.244	$\pm$	0.019	&	0.278	$\pm$	0.006	\\
2MASS J07013884-6236059	&	-0.987	$\pm$	0.007	&	3720	$\pm$	50	&	0.63	$\pm$	0.04	&	0.774	$\pm$	0.020	\\
      TYC 8602-718-1	&	-0.436	$\pm$	0.008	&	4950	$\pm$	40	&	0.854	$\pm$	0.026	&	0.824	$\pm$	0.021	\\
            HD 83096	&	0.686	$\pm$	0.007	&	6760	$\pm$	60	&	1.53	$\pm$	0.09	&	1.61	$\pm$	0.03	\\
             iot Hyi	&	0.630	$\pm$	0.010	&	6520	$\pm$	40	&	1.43	$\pm$	0.13	&	1.62	$\pm$	0.04	\\
2MASS J04082685-7844471	&	-0.964	$\pm$	0.004	&	3880	$\pm$	50	&	0.658	$\pm$	0.020	&	0.729	$\pm$	0.019	\\
2MASS J08194309-7401232	&	-1.218	$\pm$	0.017	&	3081	$\pm$	26	&	0.33	$\pm$	0.06	&	0.864	$\pm$	0.009	\\
            HD 95086	&	0.829	$\pm$	0.003	&	7573	$\pm$	9	&	1.587	$\pm$	0.024	&	1.508	$\pm$	0.007	\\
      TYC 9200-446-1	&	-0.141	$\pm$	0.006	&	5570	$\pm$	40	&	0.97	$\pm$	0.04	&	0.913	$\pm$	0.017	\\
\end{tabular}
\caption{Stellar properties (luminosity, effective temperature, mass and radius) inferred by the code used in \autoref{scarage}.}
\label{tsprops}
\end{table*}

\bsp

\end{document}